\documentclass[a4paper,11pt]{article}
\usepackage{pos}

\def\beqn{\begin{eqnarray}} \def\eeqn{\end{eqnarray}}
\def\beq{\begin{equation}} \def\eeq{\end{equation}}

\title{%%% for preprint version only:
\vspace*{-1.5cm}
\begin{minipage}{\textwidth}
{\normalfont\small 
\hspace{\fill} September 2024
}\\
\end{minipage}\\[60pt]
  From Feynman integrals to quantum algorithms: the Loop-Tree Duality connection}

\ShortTitle{From Feynman integrals to quantum algorithms: the LTD connection}

\author*[a]{German F. R. Sborlini}

\affiliation[a]{Departamento de F\'isica Fundamental e IUFFyM, Universidad de Salamanca, 37008 Salamanca, Spain.}

\emailAdd{german.sborlini@usal.es}

\abstract{In the context of high-energy particle physics, a reliable theory-experiment confrontation requires precise theoretical predictions. This translates into accessing higher-perturbative orders, and when we pursue this objective, we inevitably face the presence of complicated multi-loop Feynman integrals. There are serious bottlenecks to compute them with classical tools: the time to explore novel technologies has arrived. In this work, we study the implementation of quantum algorithms to optimize the integrands of scattering amplitudes. Our approach relies on the manifestly causal Loop-Tree Duality (LTD), which re-casts the loop integrand into phase-space integrals and avoids spurious non-physical singularities. Then, we codify this information in such a way that a quantum computer can understand the problem, and build Hamiltonians whose ground state are directly related to the causal representation. Promising results for generic families of multi-loop topologies are presented.}

\FullConference{42nd International Conference on High Energy Physics (ICHEP2024)\\
18-24 July 2024\\
Prague, Czech Republic\\}

\begin{document}
\maketitle

%%%%%%%%%%%%%%%%%%%%%%%%%%%%%%%%%%%%%%%%%%%%%%%%%%%%%%%%%%%%%%%%%%%%%%%%%%%%%%%%%%%%%%%%%%%%%%%%%
\section{Motivation}
\label{sec:Motivation}
%%%%%%%%%%%%%%%%%%%%%%%%%%%%%%%%%%%%%%%%%%%%%%%%%%%%%%%%%%%%%%%%%%%%%%%%%%%%%%%%%%%%%%%%%%%%%%%%%
High-precision calculations in perturbative quantum field theories (QFTs) require to deal with virtual amplitudes (loops) and real-radiation contributions in order to obtain finite results. Virtual amplitudes involve Feynman integrals, mathematical objects encoding the circulation of a virtual state as a quantum fluctuation of the vacuum. Most of the currently used approaches are based on the separated calculation of virtual and real amplitudes: unfortunately, these ingredients contain divergences which need to be properly cured before getting a physical prediction out from the calculation. Roughly speaking, there are three kinds of singularities: ultraviolet (UV) -present only in the Feynman integrals which are caused by the virtual states reaching very high energies-, infrared (IR) -present both in Feynman integrals and in the real-radiation contributions originated by the emission of parallel or zero-energy particles- and threshold singularities -which are due to virtual states reaching enough energy to become real particles. The last ones are integrable singularities, although they introduce serious numerical instabilities. The UV singularities are removed through the renormalization technique, whilst the cancellation of IR singularities require to add the real and virtual corrections. 

Even if dealing with higher-order calculations and curing the  aforementioned singularities are well-known topics in the literature, the efficiency of the existent methods drops very fast when we turn to multi-loop multi-particle processes. To solve this issue, several new techniques were proposed in the recent years \cite{Heinrich:2020ybq} although HEP community is still far from finding a new standard to replace the successful subtraction-based methods \cite{Frixione:1995ms,Catani:1996jh}. As a promising candidate, the Loop-Tree Duality (LTD) \cite{Catani:2008xa,deJesusAguilera-Verdugo:2021mvg} formalism offers the possibility to combine all the ingredients within a unified calculation, avoiding the proliferation of divergences in intermediate steps. Furthermore, it was recently found \cite{Verdugo:2020kzh,Aguilera-Verdugo:2020nrp,Capatti:2020ytd,Capatti:2022mly} that LTD naturally leads to very compact expressions preserving causality and only physical singularities within the loop: this is the so-called \emph{causal representation} of multi-loop multi-leg scattering amplitudes.

On top of the development of new theoretical techniques to optimize the calculations, HEP community is also in the search of novel technologies to implement the simulations in more efficient hardware and quantum computing (QC) is a potential candidate \cite{Delgado:2022tpc,diMeglio:2023nsa}. In this direction, applications of quantum algorithms (QA) to scattering amplitudes \cite{Agliardi:2022ghn,Chawdhry:2023jks}, jet clustering \cite{deLejarza:2022bwc}, Feynman integral calculations \cite{deLejarza:2023qxk,deLejarza:2024pgk} and parton shower simulations \cite{Bepari:2021kwv} have started to be explored very recently, and the list keeps on growing \cite{Rodrigo:2024say}.  

In this article, we briefly explore the fusion of LTD and QA with the purpose of identifying more efficient strategies to handle multi-loop multi-leg scattering amplitudes. We start by reviewing the basis of LTD, causal LTD and their geometrical interpretation in Sec. \ref{sec:LTD}. Then, we explain how the causal representation can be bootstrapped using QAs in Sec. \ref{sec:QuantumAlgorithms}, based on the ideas developed in Refs. \cite{Ramirez-Uribe:2021ubp,Clemente:2022nll,Ramirez-Uribe:2024wua}. Finally, we present the conclusions and future research lines in Sec. \ref{sec:conclusions}.

%20240906: OK!

%%%%%%%%%%%%%%%%%%%%%%%%%%%%%%%%%%%%%%%%%%%%%%%%%%%%%%%%%%%%%%%%%%%%%%%%%%%%%%%%%%%%%%%%%%%%%%%%%
\section{Introduction to Causal Loop-Tree Duality}
\label{sec:LTD}
%%%%%%%%%%%%%%%%%%%%%%%%%%%%%%%%%%%%%%%%%%%%%%%%%%%%%%%%%%%%%%%%%%%%%%%%%%%%%%%%%%%%%%%%%%%%%%%%%
The Loop-Tree Duality (LTD) theorem is based on the application of Cauchy residue theorem (CRT) to remove one degree-of-freedom per loop integral. Since loop integrals in QFT are usually defined on Minkowski space-time, LTD looks to remove the time or energy component in order to render the integration domain Euclidean. The main motivation for working in Euclidean space is that the infrared and soft singularities are confined in a compact region and can be directly mapped into the real-emission terms to achieve a final finite result.

In the original formulation of LTD \cite{Catani:2008xa}, Feynman propagators were replaced by the so-called \emph{dual} propagators whose prescription encoded the effect of integrating out the energy component. Still, in that framework, obtaining the dual representation for multi-loop multi-leg amplitudes could be cumbersome, specially due to the non-trivial combinations of Feynman and dual propagators appearing at higher orders. Thus, a novel approach to LTD was developed: the nested residue strategy \cite{Verdugo:2020kzh,Aguilera-Verdugo:2020nrp}. The idea is to iterate the application of CRT to each loop, which is equivalent to cut one internal line per loop. Schematically, given an arbitrary multi-loop multi-leg scattering amplitude ${\cal A}$, the result of the evaluation of the $r$-th nested residue is
\beq
{\cal A}_D(1\ldots r;r+1 \ldots) = -2\pi \imath \, \sum_{i_r \in r} {\rm Res}\left({\cal A}_D(1\ldots r-1;r \ldots),{\rm Im}(\eta \cdot q_{i_r})<0 \right) \, ,
\label{eq:IteratedRes}
\eeq
where the sum is performed over all the poles associated to the lines included in the $r$-th set of momenta that lie inside the integration contour\footnote{More details about this formulation can be found in Refs. \cite{Verdugo:2020kzh,Aguilera-Verdugo:2020nrp,Aguilera-Verdugo:2020kzc}.}. Surprisingly, it was found that several contributions (i.e. possible combinations of pole evaluations) vanish in the final expression: only those terms that can be mapped to a tree-level-like topology survive.

When all the contributions associated to the different tree-level-like topologies generated by the nested residues are put together, an even more powerful simplification occurs: the resulting expression only contains physical singularities and can be expressed in terms of \emph{causal propagators} \cite{Aguilera-Verdugo:2020kzc,Ramirez-Uribe:2020hes}. The so-called causal representation of an $L$-loop scattering amplitude is given by
\beq
{\cal A} = \sum_{\sigma \in \Sigma} \, \int_{\vec{\ell}_1 \ldots \vec{\ell}_L} \, \frac{{\cal N}_{\sigma}}{q_{1,0}^{(+)} \ldots q_{L+k,0}^{(+)}} \, \times \, \prod_{i=1}^k \frac{-1}{\lambda_{\sigma(i)}} \ + (\sigma \longleftrightarrow \bar{\sigma}) \, ,
\label{eq:CausalRep}
\eeq
where $q_{i,0}^{(+)}$ is the positive on-shell energy associated to $i$-th internal line, $k=V-1$ is the order of the topology, $V$ the number of interaction vertices and $\lambda_j = \sum_l q_{l,0}^{(+)} \pm p_j$ denote the causal thresholds. $\sigma$ is a combination of $k$ compatible thresholds, and the set $\Sigma$ contains all the allowed causal entangled thresholds, which define the structure of singularities of the corresponding scattering amplitude. It turns out that $\Sigma$ can be obtained based on geometrical rules: in fact, the causal representation can be bootstrapped from the collection of all the directed acyclic graphs (DAG) associated to the Feynman graphs describing the scattering amplitude \cite{Sborlini:2021owe,Ramirez-Uribe:2021ubp}.

%20240911: OK!!

%%%%%%%%%%%%%%%%%%%%%%%%%%%%%%%%%%%%%%%%%%%%%%%%%%%%%%%%%%%%%%%%%%%%%%%%%%%%%%%%%%%%%%%%%%%%%%%%%
\section{Feynman integrals and quantum algorithms}
\label{sec:QuantumAlgorithms}
%%%%%%%%%%%%%%%%%%%%%%%%%%%%%%%%%%%%%%%%%%%%%%%%%%%%%%%%%%%%%%%%%%%%%%%%%%%%%%%%%%%%%%%%%%%%%%%%%
The connection between DAGs and the causal representation is the key observation to tackle multi-loop Feynman integrals with quantum algorithms (QA). In concrete, we reformulate the original problem: instead of dealing with the amplitude itself, we consider the underlying \emph{reduced} graph and look for all the DAGs. Thus, we deploy QAs capable of efficiently identifying all the DAGs of a given graph. In Ref. \cite{Ramirez-Uribe:2021ubp}, we apply a Grover-based algorithm to detect DAGs using binary clauses to codify the acyclic condition. Even if the performance on simulators is excellent, this kind of algorithms faces serious challenges in noise intermediate-scale quantum (NISQ) devices.

Thus, we explore another approach based on Variational Quantum Eigensolvers (VQE), which are better suited for NISQ computers. Roughly speaking, VQE is a minimization hybrid algorithm that aims to identify the ground state of Hamiltonians. In this case, we codify the acyclic condition through a Hamiltonian created from the adjacency matrix $A$ of the reduced Feynman graph. In concrete, if we promote the adjacency matrix to an operator $\hat{A}$ acting on the space of vertices $V$ and edges $E$, then for a given graph $G=(E,V)$ we define
\beq
\hat{H_G} = \sum_{n=1}^V \, {\rm Tr}_V \left(\hat{A}^n\right) \, .
\label{eq:Hamiltonian}
\eeq
This Hamiltonian acts on the space of edges, being the direction of each edge encoded in a single qubit, and its ground state contains all the bit-strings associated to DAGs. In Ref. \cite{Clemente:2022nll}, we implemented a VQE to solve this Hamiltonian for a set of multi-loop topologies. For the simplest cases, we found that the standard VQE is able to successfully identify almost all the DAGs. 

%%%%%%%%%%%%%%%%%%%%%%%%%%%%%%
\begin{figure}[htb]
\begin{center}
\includegraphics[scale=0.41]{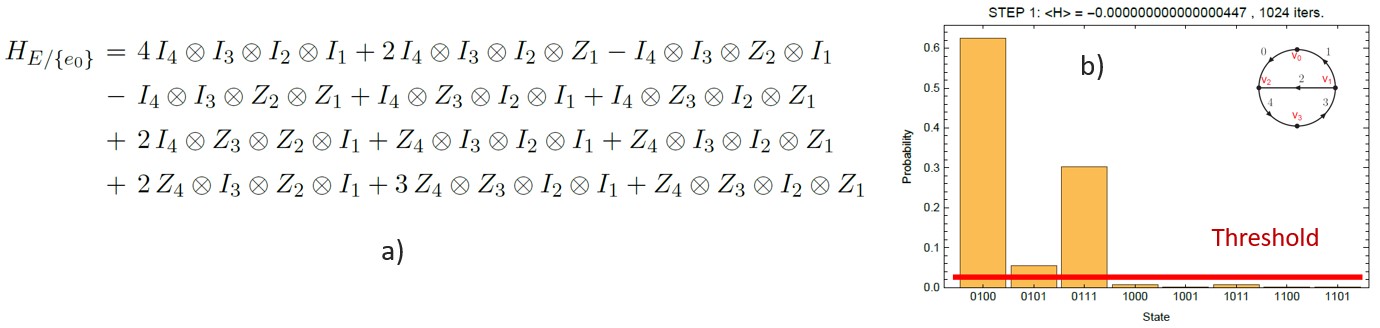}
\caption{(a) Explicit form of the Hamiltonian in Eq. (\ref{eq:Hamiltonian}) for a 2-loop 5-edge topology. (b) Output of the single-run VQE. The red line indicates a threshold to collect elements that belong to the ground state.}
\label{fig:Figura1}
\end{center}
\end{figure}
%%%%%%%%%%%%%%%%%%%%%%%%%%%%%

The number of DAGs grows very fast with the number of edges $E$, which implies that the ground state of $\hat{H_G}$ is highly degenerated. In Fig. \ref{fig:Figura1}, we show an example of a 2-loop 5-edge topology and the output after the single-execution of VQE: we only tag 3 out of 18 DAGs. Thus, we improved the algorithm implementing a multi-run VQE. The idea is to run several times the VQE, collecting in each step some DAGs and introducing penalisation terms in the Hamiltonians of the subsequent executions to avoid double-counting. With this strategy, we reach success rates of ${\cal O}(90 \, \%)$ for topologies with hundreds of DAGs. Furthermore, by properly choosing the starting point of each run, we manage to improve the speed of the convergence and also (partially) avoid getting stuck in false minima (which makes the algorithm stop and prevents the identification of all the solutions). More details can be found in Ref. \cite{Clemente:2022nll}.

%20240911: OK!!

%%%%%%%%%%%%%%%%%%%%%%%%%%%%%%%%%%%%%%%%%%%%%%%%%%%%%%%%%%%%%%%%%%%%%%%%%%%%%%%%%%%%%%%%%%%%%%%%%%%%%%%%%%%%%%%%%%%%%%%%%%%
\section{Conclusions}
\label{sec:conclusions}
%%%%%%%%%%%%%%%%%%%%%%%%%%%%%%%%%%%%%%%%%%%%%%%%%%%%%%%%%%%%%%%%%%%%%%%%%%%%%%%%%%%%%%%%%%%%%%%%%%%%%%%%%%%%%%%%%%%%%%%%%%%
In this article, we briefly explain the potential advantages of the Loop-Tree Duality (LTD) and its manifestly causal representation for multi-loop amplitude calculations. On one side, this representation is defined on Euclidean space and contains only physical singularities. On the other, causality naturally leads to a geometrical connection that can be further exploited to re-formulate the problem in such a way that quantum computers can handle it. In this direction, we commented on the application of VQE and multi-run VQE \cite{Clemente:2022nll} to obtain all the directed acyclic graphs (DAG) required to bootstrap the causal representation of multi-loop topologies.

The causal representation of multi-loop amplitudes offers new directions to explore the calculation of physical observables and cross-sections. We have recently developed a formalism \cite{Ramirez-Uribe:2024rjg,LTD:2024yrb} to compute higher-order corrections to full cross-sections starting from the causal representation of multi-loop vacuum diagrams. Since vacuum diagrams were successfully studied with our quantum algorithms \cite{Ramirez-Uribe:2021ubp,Clemente:2022nll}, in the future, we expect to be able to perform a full cross-section calculation with a quantum computer, being the LTD the connection between loop amplitudes and quantum algorithms.

%20240910: OK!

%%%%%%%%%%%%%%%%%%%%%%%%%%%%%%%%%%%%%%%%%%%%%%%%%%%%%%%%%%%%%%%%%%%%%%%%%%%%%%%%%%%%%%%%%%%%%%%%%%%%%%%%%%%%%%%%%%%%%%%%%%%
\subsection*{Acknowledgments}
We would like to thank K. Pyretzidis for reading this manuscript. This work is supported by the Spanish Government (Agencia Estatal de Investigaci\'on MCIN /AEI/10.13039/501100011033) Grants No. PID2022-141910NB-I00, Generalitat Valenciana Grant No. PROMETEO/2021/071 and H2020-MSCA-COFUND USAL4EXCELLENCE-PROOPI-391 project under Grant Agreement No 101034371.

%%%%%%%%%%%%%%%%%%%%%%%%%%%%%%%%%%%%%%%%%%%%%%%%%%%%%%%%%%%%%%%%%%%%%%%  

%\bibliographystyle{JHEP}
%\bibliography{refs}

\providecommand{\href}[2]{#2}\begingroup\raggedright\endgroup

\end{document}